\documentclass[pra,twocolumn,showpacs,amsmath,amssymb]{revtex4}
\usepackage{amssymb}
\usepackage{dcolumn}
\usepackage{bm}
\usepackage{graphicx}
\usepackage{amsmath}
\usepackage{CJK}
\begin{document}
\begin{CJK*}{GB}{gbsn}
\title{Nonlinear dissipative dynamics of a two-component atomic condensate coupling with a continuum}

\author{Zhong Honghua$^{1,2}$(钟宏华), Xie Qiongtao$^{2,3}$(谢琼涛), Xu Jun$^{2}$(徐军), Hai Wenhua$^{3}$(海文华)}

\author{Li Chaohong$^{2}$(李朝红)}

\affiliation{$^{1}$Department of Physics, Jishou University, Jishou 416000, China}
\affiliation{$^{2}$State Key Laboratory of Optoelectronic Materials and Technologies, School of Physics and Engineering, Sun Yat-Sen University, Guangzhou 510275, China}

\affiliation{$^{3}$Department of Physics and Key Laboratory of Low-Dimensional Quantum Structure and Quantum Control of Ministry of Education, Hunan Normal University, Changsha 410081, China}

\date{\today}

\begin{abstract}
We investigate the nonlinear dissipative coherence bifurcation and population dynamics of a two-component
atomic Bose-Einstein condensate coupling with a continuum. The coupling between the
two-component condensates and the continuum brings effective dissipations to the two-component
condensates. The steady states and the coherence bifurcation depend on both dissipation and the
nonlinear interaction between condensed atoms. The coherence among condensed atoms may be
even enhanced by the effective dissipations. The combination of dissipation and nonlinearity allows
one to control the switching between different self-trapped states or the switching between a self-trapped
state and a non-self-trapped state.

\end{abstract}

\maketitle
\end{CJK*}
\section{Introduction}

Since the experimental realization of the two-component condensates, such as Bose-Einstein condensates (BECs) of $^{87}Rb$ in two different hyperfine levels~\cite{hall,ensher,wieman,harber,bongs,zibold}, the dynamical features of two-component condensates have been widely investigated in recent years. Within the mean-field description, several macroscopic quantum behaviors and
interesting dynamical properties~ \cite{liu,cooper,walser,gao1,Santos,Tsubota,hai,Weyrauch,wliu,Zurek,gao,cirac,ueda1,Bezett,kivshar,Lee2012} have been explored. Within the quantum-field description, several many-body quantum phenomena~ \cite{qtxie,shen,Lee2006,ng,kishi,Lee20062,jaksch,Lee2008,csire,Lee2009}, such as quantum self-trapping~\cite{shen} and spontaneous symmetry breaking~\cite{Lee2009}, have been predicted. Some of these dynamical properties have also been observed in experiments~\cite{hall,sadler}.

In real experiments, due to the atomic loss induced by inelastic collisions and the coexistence of condensed and non-condensed atoms~\cite{sols,donley,pethick}, the atomic condensates are not absolutely closed systems. The dissipative processes can be effectively described by non-Hermitian (NH) Hamiltonians~\cite{adhi,ueda,graefe,korsch0,kohler}. The dissipation-induced effects in a NH Bose-Hubbard dimer with a complex on-site energy \cite{korsch2,korsch,korsch3,hiller} or a complex coupling term \cite{zhong} have been studied extensively. If a system has an intrinsic mechanism balancing the dissipations, the dissipation can lead to a constructive effect, such as the enhanced self-trapping \cite{korsch,korsch3} and the inhibited losses of atoms \cite{konotop}. Moreover, the dissipation-induced coherence in an open two-mode BEC system have been studied by the master equation method \cite{swimber, guqiang}.

In single-particle systems, the interaction between a two-state model and a continuum is a paradigm to understand the quasi-stationary states~\cite{seby,makhmet}. Up to now, there is still not any study on the coupling between a many-body two-state system with continuum. The experimental realization of two-component BEC provides a new possibility for exploring many-body quantum phenomena of atoms coupled with continuum, such as many-body coherence, dissipative dynamics and population transition. Therefore, the study  of two-component condensates
coupling with continua will provide a benchmark
for understanding dissipative many-body quantum
systems.

In this article, we investigate the many-body quantum coherence and population dynamics in a two-component condensate coupled with a continuum. Our system can
be described by an effective NH Bose-Hubbard dimer
with complex diagonal and off-diagonal elements, which
are induced by the continuum. In the mean-field theory, the system obeys a two-mode NH Gross-Pitaevskii Hamiltonian, in which the nonlinear terms describe the atom-atom interactions. We find that the combination of dissipation and nonlinearity may induce different steady states and modify the coherence bifurcation. By tuning the nonlinearity and dissipation strength, it is possible to observe the coherence enhancement. Particularly, different relative dissipation strengths between two hyperfine levels will drive the two-component condensates into different stable states. In the time evolution, the system can jump from one stable state to another stable state and the switching time  depends on the nonlinearity strength. The combination of dissipation and nonlinearity can be used to manipulate the steady behaviors, such as, controlling the transition between two self-trapping states or between a self-trapped state and a non-self-trapped state. Therefore, our results provide an alternative route for manipulating the many-body coherence and population dynamics by controlling and  utilizing dissipation and nonlinearity.

The structure of this article is as following. In section II, we give the non-Hermitian Hamiltonian and the equations of motion. In section III, we analyze that how dissipation and nonlinearity affect the bifurcation of coherence. In section IV, we show various nonlinear dissipative dynamics under different conditions. In the last section, we briefly summarize our results.

\section{Non-Hermitian two-mode Bose-Hubbard model}

We consider the interaction between a two-component condensate, whose atoms occupy two different hyperfine levels, and a continuum. The two hyperfine levels are coupled by a Raman laser, and interact with a
common continuum by external fields \cite{seby,fano,anglin}. With the well-known  single-mode approximation, and by eliminating atoms in continuum
\cite{seby,makhmet}, the system obeys the Hamiltonian
\begin{eqnarray}
\hat{H}=&&\left(\Omega -i\sqrt{\Gamma_{1}\Gamma_{2}}/2\right) \left(b_{1}^\dagger b_{2}+b_{2}^\dagger b_{1}\right) +U_{12}b_{1}^{\dagger}b_{2}^{\dagger}b_{2}b_{1} \nonumber
\\ &&+\sum_{j=1,2} \left[\left(\varepsilon_{j} -i\frac{\Gamma_{j}}{2}\right)b_{j}^{\dagger}b_{j}
+\frac{U_{jj}}{2}b_{j}^{\dagger}b_{j}^{\dagger} b_{j}b_{j}\right].
\end{eqnarray}
Here $b_{j}^\dagger$ ($b_{j}$) are the bosonic creation (annihilation) operators for atoms occupying the $j$-th hyperfine level. In the complex coupling term, $\Omega-i\sqrt{\Gamma_{1}\Gamma_{2}}/2$, $\Omega$ denotes the direct coupling between the two hyperfine levels and $i\sqrt{\Gamma_{1}\Gamma_{2}}/2$ describes the effective coupling between them via the continuum~\cite{seby,makhmet}. The inter-component interaction $U_{jj}$ and the intra-component interaction $U_{12}$ can be controlled by adjusting the atomic s-wave scattering lengthes via Feshbach resonance techniques. In the complex energies, $\varepsilon_{j}-i\Gamma_{j}/2$, the imaginary parts $\Gamma_{j}$ denote dissipation rates for atoms in the $j$-th hyperfine level due to the interaction with the continuum~\cite{seby,makhmet}. For the case of $\Gamma_{2}=0$, the system (1) is just the model studied in~\cite{korsch,korsch3}. For the case of no atom-atom interaction, $U_{jj}=U_{12}=0$, the system (1) is reduced to the single-atom system~\cite{seby,makhmet}. To simplify, we set $\hbar=1$ and adopt $\Omega$ as a unit to rescale the other parameters $\varepsilon_{j}$, $\Gamma_{j}$, $U_{12}$ and $U_{jj}$.

To investigate the dynamics of the system (1) in the Bloch representation, we introduce the angular momentum operators,
\begin{eqnarray}
L_{x}&=&\frac{1}{2}(b_{1}^\dagger b_{2}+b_{1}b_{2}^\dagger),\nonumber
\\ L_{y}&=&\frac{1}{2i}(b_{2}^\dagger b_{1}- b_{1}^\dagger b_{2}),
\nonumber
\\ L_{z}&=&\frac{1}{2}(b_{2}^\dagger
b_{2}-b_{1}^\dagger b_{1}),
\end{eqnarray}
whose Casimir invariant is $L^{2}=(N/2)(N/2+1)$. Here $N=b_{1}^\dagger b_{1} +b_{2}^\dagger b_{2}$ is the total number operator. With these angular momentum operators, by omitting the real constant terms $O(N)$ and $O(N^2)$ \cite{korsch,korsch3}, the Hamiltonian (1) can be written as
\begin{eqnarray}
\hat{H}=2(\Omega-i\Lambda)L_{x}+G L_{z}^{2} +
(\delta-i\gamma)L_{z}-i\Upsilon N,
\end{eqnarray}
where $\Lambda=\sqrt{\Gamma_{1}\Gamma_{2}}/2$, $G=(U_{11}+U_{22}-2U_{12})/2$, $\delta=\varepsilon_{2}-\varepsilon_{1} +(U_{22}-U_{11})(N-1)/2$, $\gamma=(\Gamma_{2}-\Gamma_{1})/2$ and
$\Upsilon=(\Gamma_{1}+\Gamma_{2})/4$.

To investigate the mean-field dynamics, introducing the $SU(2)$ coherent states
\begin{eqnarray}
|x_{1},x_{2}\rangle=\frac{1}{\sqrt{N!}}(x_{1}b_{1}^{\dag}+x_{2}b_{2}^{\dag})^{N}|0,0\rangle,
\end{eqnarray}
with two complex coefficients $x_{j}$, the  mean-field Bloch vectors $s_{k}=\langle L_{k}\rangle/N$, $k=x,y,z$, read as
\begin{eqnarray}
s_{x}=\frac{\langle x_{1},x_{2}|L_{x}|x_{1},x_{2}\rangle}{\langle
x_{1},x_{2}|x_{1},x_{2}\rangle N}&=&
\frac{(x_{1}^{\ast}x_{2}+x_{1}x_{2}^{\ast})}{2n}, \nonumber
\\ s_{y}= \frac{\langle x_{1},x_{2}|L_{y}|x_{1},x_{2}\rangle}{\langle
x_{1},x_{2}|x_{1},x_{2}\rangle N}&=&
\frac{(x_{2}^{\ast}x_{1}-x_{1}^{\ast}x_{2})}{2 i n}, \nonumber
\\ s_{z}=\frac{\langle x_{1},x_{2}|L_{z}|x_{1},x_{2}\rangle}{\langle
x_{1},x_{2}|x_{1},x_{2}\rangle N}&=&
\frac{(x_{2}^{\ast}x_{2}-x_{1}^{\ast}x_{1})}{2n}.
\end{eqnarray}
Here, $n=|x_{1}|^2+|x_{2}|^2$ is the norm. By using the method developed in Refs.~\cite{korsch,korsch3}, the equations of motion for the Bloch vector are given as
\begin{eqnarray} \label{nhbequation}
\dot{s_{x}}&=&-(\delta+2Cs_{z})s_{y}+2\gamma s_{z}s_{x}-
\Lambda(1-4s_{x}^{2}), \nonumber
\\ \dot{s_{y}}&=&(\delta+2Cs_{z})s_{x}-2\Omega s_{z}+
4\Lambda s_{x}s_{y}+2\gamma s_{z}s_{y}, \nonumber
\\ \dot{s_{z}}&=&2\Omega s_{y}+
4\Lambda s_{z}s_{x}-\gamma(1-4s_{z}^{2})/2.
\end{eqnarray}
To obtain these equation, we have taken the semiclassical limit $N\rightarrow\infty$ with the mean-field interaction strength $C=NG$ kept unchanged. Therefore, $s^{2}=s_{x}^{2}+s_{y}^{2}+s_{z}^{2}=1/4$ is a constant and the dynamics are regular and confined onto the Bloch sphere. However, the total probability $n$ decays as
\begin{eqnarray}
\dot{n}&=&-2(2\Lambda s_{x}+\gamma s_{z}+\Upsilon)n.
\end{eqnarray}
The Bloch equation (6) also can equivalent to a generalized NH nonlinear Schr\"{o}dinger equation. Introducing the unnormalized complex numbers $\psi_{j}$, which are associated with the coefficients $x_{j}$ of the many-particle coherent state~\cite{korsch}, the Bloch equation (\ref{nhbequation}) corresponds to the NH nonlinear Schr\"{o}dinger equation
\begin{eqnarray} \label{twomodeequation}
&&i\frac{d}{dt}\left( \begin{array}{c}
\psi_{1} \\
\psi_{2} \\
\end{array}
 \right)  \nonumber \\
&&=\left(
     \begin{array}{cc}
      -\frac{\delta}{2}-
C\kappa-i(\Upsilon-\frac{\gamma}{2}) & \Omega-i\Lambda \\
     \Omega-i\Lambda & \frac{\delta}{2}+
C\kappa-i(\Upsilon+\frac{\gamma}{2}) \\
     \end{array}
   \right) \left( \begin{array}{c}
\psi_{1} \\
\psi_{2} \\
\end{array}
 \right), \nonumber \\
\end{eqnarray}
with
$\kappa=(|\psi_{2}|^{2}-|\psi_{1}|^{2}) /(|\psi_{1}|^{2}+|\psi_{2}|^{2})$. Similar NH nonlinear Schr\"{o}dinger equations have been proposed to describe open atomic BECs~\cite{korsch,korsch3,xyi} and double-channel waveguide with gain and
loss recently~\cite{lilu,xluo}.

\section{Coherence bifurcation under dissipation and nonlinearity}

In this section, we investigate the combined effects of the dissipation and nonlinearity on  the phase coherence between two modes. Here we use the contrast in interference experiments to measure  the phase coherence. The contrast is defined as ~\cite{swimber,guqiang}
\begin{eqnarray}
\alpha &=&\frac{2\mid\langle b_{1}^\dagger b_{2}\rangle\mid}{\langle b_{1}^\dagger b_{1}+b_{2}^\dagger b_{2}\rangle}
=\sqrt{s_{x}^{2}+s_{y}^{2}}.
\end{eqnarray}
In this work, we mainly focus on the coherence  of the steady states (the fixed points of the equations of motion). These steady states can be calculated numerically via Newton flow method from $\dot{\vec{s}}=0$~\cite{chen}. For the case of $\delta=0$, we have
\begin{eqnarray} \label{fixpointequation}
s^{0}_{x}&=&\frac{s^{0}_{z}[Q(2\Lambda Cs^{0}_{z}-\gamma \Omega)+4\Omega(\Lambda^{2}+\Omega^{2})]}{2[Q\Lambda \Omega+2Cs^{0}_{z}(4\Lambda^{2} (s^{0}_{z})^{2}+\Omega^{2})]}, \nonumber \\
s^{0}_{y}&=&\frac{Q-8\Lambda s^{0}_{z} s^{0}_{x}}{4\Omega},
\end{eqnarray}
where $Q=\gamma[1-4(s^{0}_{z})^2]$ and $s^{0}_{k}$ denote the corresponding fixed points. By using Eq.~(\ref{fixpointequation}) and the normalization condition $s^{2}=(s_{x}^0)^{2}+(s_{y}^0)^{2}+(s_{z}^0)^{2}=1/4$, we determine the values of $s^{0}_{z}$ and then $s_x^0$ and $s_y^0$ from Eq.~(\ref{fixpointequation}). A key issue is the stability of the fixed points. To do this, we linearize Eq.~(\ref{nhbequation}) around the fixed points by substituting
\begin{eqnarray}
s_{x}=s^{0}_{x}+\Delta s_{x}, s_{y}=s^{0}_{y}+\Delta s_{y},
s_{z}=s^{0}_{z}+\Delta s_{z},
\end{eqnarray}
with a small deviation $(\Delta s_{x}$, $\Delta s_{y}$, $\Delta s_{z})$ into Eq.~(\ref{nhbequation}), thereby one can obtain a linearized equation for this small deviation~\cite{swimber,qiongtao,vardi}
\begin{eqnarray} \label{linearizeequation}
\frac{d}{dt}\left( \begin{array}{c}
\Delta s_{x} \\
\Delta s_{y} \\
\Delta s_{z} \\
\end{array}
\right)=M
 \left( \begin{array}{c}
\Delta s_{x} \\
\Delta s_{y} \\
\Delta s_{z} \\
\end{array}
\right),
\end{eqnarray}
with the coefficient matrix
\begin{eqnarray} \label{matixequation}
M=\left(\begin{array}{ccc}
2\gamma s^{0}_{z}+8\Lambda s^{0}_{x} & -2Cs^{0}_{z} & 2\gamma s^{0}_{x}-2C s^{0}_{y} \\
2C s^{0}_{z}+4\Lambda s^{0}_{y} & 4\Lambda s^{0}_{x}+2\gamma s^{0}_{z} & 2C s^{0}_{x}-2\Omega+2\gamma s^{0}_{y}  \\
4\Lambda s^{0}_{z} & 2\Omega & 4\Lambda s^{0}_{x}+4\gamma s^{0}_{z}\\
\end{array} \right). \nonumber \\
\end{eqnarray}
The fixed points are linearly stable if and only if there is no eigenvalues with a positive real part for the coefficient matrix $M$. The eigenvalues of the coefficient matrix can be obtained numerically.

\begin{figure}[htp] \center
\includegraphics[width=1.0\columnwidth]{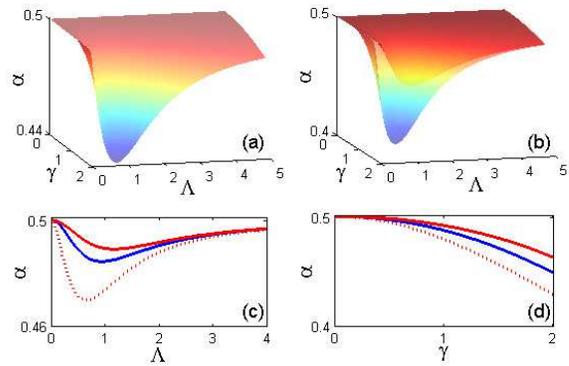}
\caption{(Color online) Dependence of the steady contrast $\alpha$ on the relative dissipation strength $\gamma$ and the imaginary coupling strength $\Lambda$. The nonlinearity $C$ is chosen as (a) 0 and (b) 1. (c) and (d) are sections of (a) and (b) for $\gamma=1$ and $\Lambda=1$, respectively. The red solid (dotted) lines correspond to the stable (unstable) steady states for the nonlinear case, and the blue solid lines are steady states for the linear case. The other parameters are given as $\Omega=1$ and $\delta=0$.}
\end{figure}

We show our numerical results for the steady contrast $\alpha$ in Fig.~1. The dependence of $\alpha$ on the relative dissipation strength $\gamma$ and the imaginary coupling strength $\Lambda$ are shown for two different cases: (a) the linear case $C=0$ and (b) the nonlinear case $C=1$. In Figs.~1(c) and 1(d), we plot the sections of Figs.~1(a) and 1(b) for $\gamma=1$ and $\Lambda=1$, respectively. The other parameters are given as $\Omega=1$ and $\delta=0$. The numerical results indicate that there exist two fixed points for the nonlinear case of $C=1$ and while there only exists one fixed point for the linear case of $C=0$. By implementing the linear stability analysis, we find that only one of two fixed points for the nonlinear case is stable, marked by the solid lines (see Figs.~1(c) and 1(d)). In particular, we observe that the steady contrast $\alpha$ for the nonlinear case is larger than the one for the linear case. This means that the coherence can be enhanced by the combination of the dissipation and the nonlinearity. In addition, we find that the coherence may be enhanced by the imaginary coupling strength $\Lambda$, see Fig.~1(c). As $\Lambda$ increases, the coherence slowly decreases first and then increases. This is similar to the dissipation induced coherence in previous works~\cite{swimber,guqiang}.

\begin{figure}[htp] \center
\includegraphics[width=1.0\columnwidth]{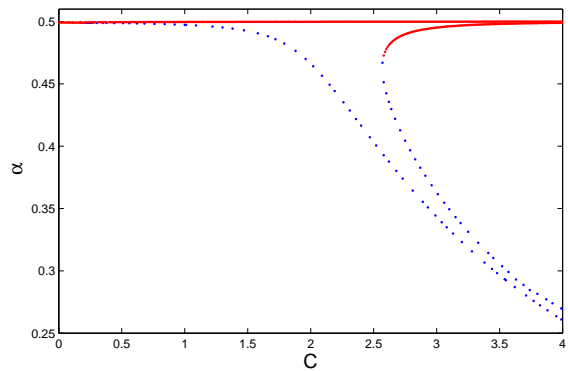}
\caption{(Color online) The steady contrast $\alpha$ versus the nonlinearity C for $\Omega=1$, $\gamma=0.4$ and $\Lambda=0.3$. Here the red  solid lines correspond to the stable fixed points, and the blue dotted lines correspond to the unstable fixed points.}
\end{figure}

In Fig. 2, we show how the nonlinearity $C$ affects the contrast $\alpha$. For weak nonlinearity, the contrast $\alpha$ has two different values corresponding to one stable and one unstable fixed points. As the nonlinearity increases, a bifurcation appears at a certain critical value of the nonlinearity. This bifurcation of the contrast $\alpha$ is associated with the occurrence of one stable and one unstable points beyond the critical nonlinearity. Here the red solid lines denote the stable fixed points and the blue dotted lines represent the unstable fixed points.

\section{Nonlinear dissipative dynamics}

Based upon our understanding of the steady states in above, in this section, we analyze the mean-field dynamical behavior arising from the interplay of dissipation and nonlinearity. For the case that the dissipation strength in one of hyperfine levels is zero, i.e. $\Lambda=0$, the corresponding NH Hamiltonian is reduced to the ones in Refs.~\cite{korsch,korsch3}. The mean-field dynamics has been studied in detail for this case, in which the dissipation enhanced self-trapping states have been revealed~\cite{korsch,korsch3}. In this work, for simplicity, we focus on considering the following two cases: (A) $\gamma=0$ and (B) $\gamma\neq0$ and $\Lambda \neq 0$.

\subsection {$\gamma=0$}

Under the condition of $\Gamma_{1}=\Gamma_{2}$, the relative dissipation between the two hyperfine levels vanishes, that is $\gamma=0$. Therefore, the mean-field Bloch equations are given as
\begin{eqnarray} \label{blochequation}
\dot{s_{x}}&=&-(\delta+2Cs_{z})s_{y}- \Lambda(1-4s_{x}^{2})/2,
\nonumber
\\ \dot{s_{y}}&=&(\delta+2Cs_{z})s_{x}-2\Omega s_{z}+
2\Lambda s_{x}s_{y}, \nonumber
\\ \dot{s_{z}}&=&2\Omega s_{y}+
2\Lambda s_{z}s_{x},
\end{eqnarray}
with the total probability $n$ satisfying
\begin{eqnarray} \label{pequation}
\dot{n}&=&-\frac{\Lambda}{2} (2s_{x}+1)n.
\end{eqnarray}
In Fig. 3, we show the dynamical behavior of the Bloch equation (\ref{blochequation}) on the Bloch sphere for the Hermitian case of $\Lambda=0$ (top) and the non-Hermitian case of $\Lambda=0.4$ (bottom) with $\Omega=1$ and $\delta=0$. For the Hermitian case of $\Lambda=0$, the Bloch vectors evolve periodically on the surface of the Bloch sphere and form closed orbits dependent upon initial conditions. There are two stable centers at $s_{y}=s_{z}=0$ and $s_{x}=\pm1/2$, which are shown in Fig. 3(a). When the nonlinearity strength $C$ increases, one of the two centers becomes unstable and a saddle and two stable centers appear after a bifurcation, which are shown in Fig. 3(b). In the case of $\Lambda=0.4$ (bottom), we see a drastic modification of these patterns. On the one hand, all orbits on the surface of the Bloch sphere become non-closed. On the other hand, the system starting from different initial states may relax to the same state of $s_{y}=s_{z}=0$ and $s_{x}=-1/2$, see Figs. 3(c) and (d). Furthermore, the change of nonlinearity and dissipation only modifies the evolution paths and dissipation velocities to the final state. Therefore, in the case of $\gamma=0$, for different nonlinearities and initial atom populations, the final state will be completely definite, that is, it is an equal population state.

To better understand the mean-field dynamics, we analyze the fixed points  and their stability. The fixed points are determined by
\begin{eqnarray} \label{zequation}
&&s^{0}_{z}=\frac{\Omega s^{0}_{x}\delta}{\Omega^{2}-2C\Omega s^{0}_{x}+\Lambda^{2}(s^{0}_{x})^{2}}, \nonumber \\
&&(s^{0}_{x})^{2}+\frac{(s^{0}_{x})^{2}\delta^{2}[\Omega^{2}+\Lambda^{2}(s^{0}_{x})^{2}]}{[\Omega^{2}-2C\Omega s^{0}_{x}
+\Lambda^{2} (s^{0}_{x})^{2})]^{2}}-\frac{1}{4}=0.
\end{eqnarray}
For the case of $\delta=0$, we have two stationary states $s^{0}_{x}=\pm \frac{1}{2}$ and $s^{0}_{y}=s^{0}_{z}=0$ from Eq.~(\ref{blochequation}). However, considering Eq. (\ref{pequation}) together, we only have one stable fixed point
\begin{eqnarray}
{s}^{0}_{k}=\left(
 \begin{array}{c}
 -\frac{1}{2}  \\
 0  \\
 0 \\
 \end{array}
 \right).
\end{eqnarray}
This explains why all initial states finally decay into this stable fixed point.

\begin{figure}[htp] \center
\includegraphics[width=1.0\columnwidth]{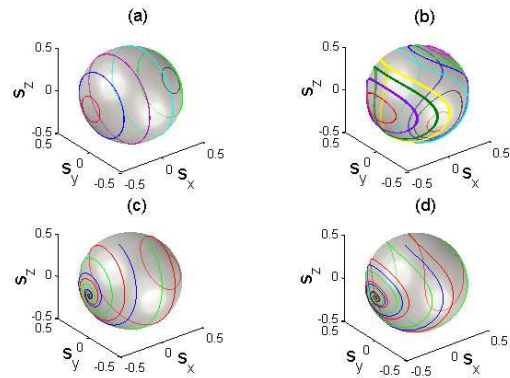}
\caption{(Color online) Mean-field dynamics on the Bloch sphere for the Hermitian ($\Lambda=0$, top) and non-Hermitian ($\Lambda=0.4$, bottom) cases for $\Omega=1$, $\delta=0$ and different values of the nonlinearity [$C=0$ (left), $2$ (right)]. Different tinctorial orbits on the Bloch sphere correspond to different initial conditions.}
\end{figure}

\subsection{$\gamma\neq0$ and $\Lambda \neq0$}

Now, we consider the general case of $\gamma\neq0$ and $\Lambda \neq0$, which obeys the Bloch equations Eq.~(6). In  Fig. 4, we show the mean-field dynamics on the Bloch sphere for the two different cases: $\gamma> \Lambda$ (top) and $\gamma < \Lambda$ (bottom) with $\Omega=1$ and $\delta=0$. Similarly, the orbits on the surface of the Bloch sphere are not closed. In the case of $\gamma > \Lambda$, it is observed that, for linear and weakly nonlinear cases, all different initial states will always evolve into the same final state, as shown in Figs. 4(a) and (b). However, for strongly nonlinear cases, different initial states will evolve into different final states. As shown in Fig. 4(c), for the case of $C=5$, two different initial states evolve into different final states. Therefore, for $\gamma > \Lambda$, the change of the nonlinearity and initial states not only can modify the evolution paths and dissipation velocities to the final state but also may
change the final state. However, for $\gamma < \Lambda$, independent upon the initial state and nonlinearity strength, the system will always evolve to the same point on the Bloch sphere, which corresponds to the same final state, while the change of nonlinearity modifies its evolution path and dissipation velocity to the final state, as shown in Figs. 4(d)-(f). After a further calculation, we find that for $\gamma > \Lambda$, the system has two stable fixed points, while $\gamma < \Lambda$, the system has only one stable fixed points. So the competition between $\gamma$ and $\Lambda$ results in different final state. It is important to note that the dissipation can be controlled by shining a laser beam onto the condensates~\cite{bloch} and relative dissipation rate $\gamma$ can also be changed at the same time. Therefore the combination of nonlinearity and dissipation can be used for controlling the dynamics.
\begin{widetext}
\begin{center}
\begin{figure}[htp] \center
\includegraphics[width=1.0\columnwidth]{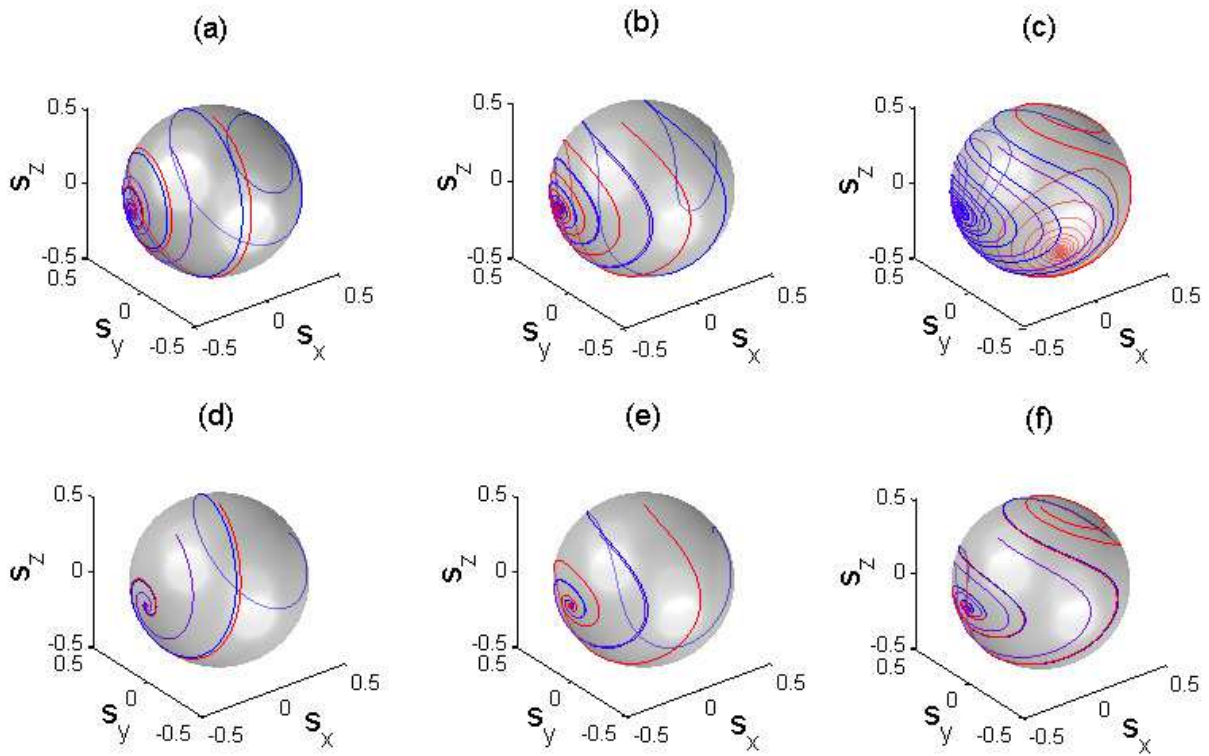}
\caption{(Color online) Mean-field dynamics on the Bloch sphere for the case $\gamma > \Lambda$ (top) and $\gamma < \Lambda$ (bottom) for $\delta=0$ and $\Omega = 1$. (a)-(c) $\gamma=0.4$, $\Lambda=0.15$; $C=0.0$ (a), $1.5$ (b), $5.0$ $(c)$. (d)-(f) $\gamma=0.05$, $\Lambda=0.324$; $C=0.0$ (d), $1.5$ (e), $5.0$ $(f)$. Here different tinctorial orbits in the Bloch sphere correspond to different initial conditions with $s(0)=[0,0,0.5]$ (red), $[0.48,0,0.14]$ (blue), $[0.3,0,0.4]$ (purple).}
\end{figure}
\end{center}
\end{widetext}

\begin{figure}[htp] \center
\includegraphics[width=3.5in]{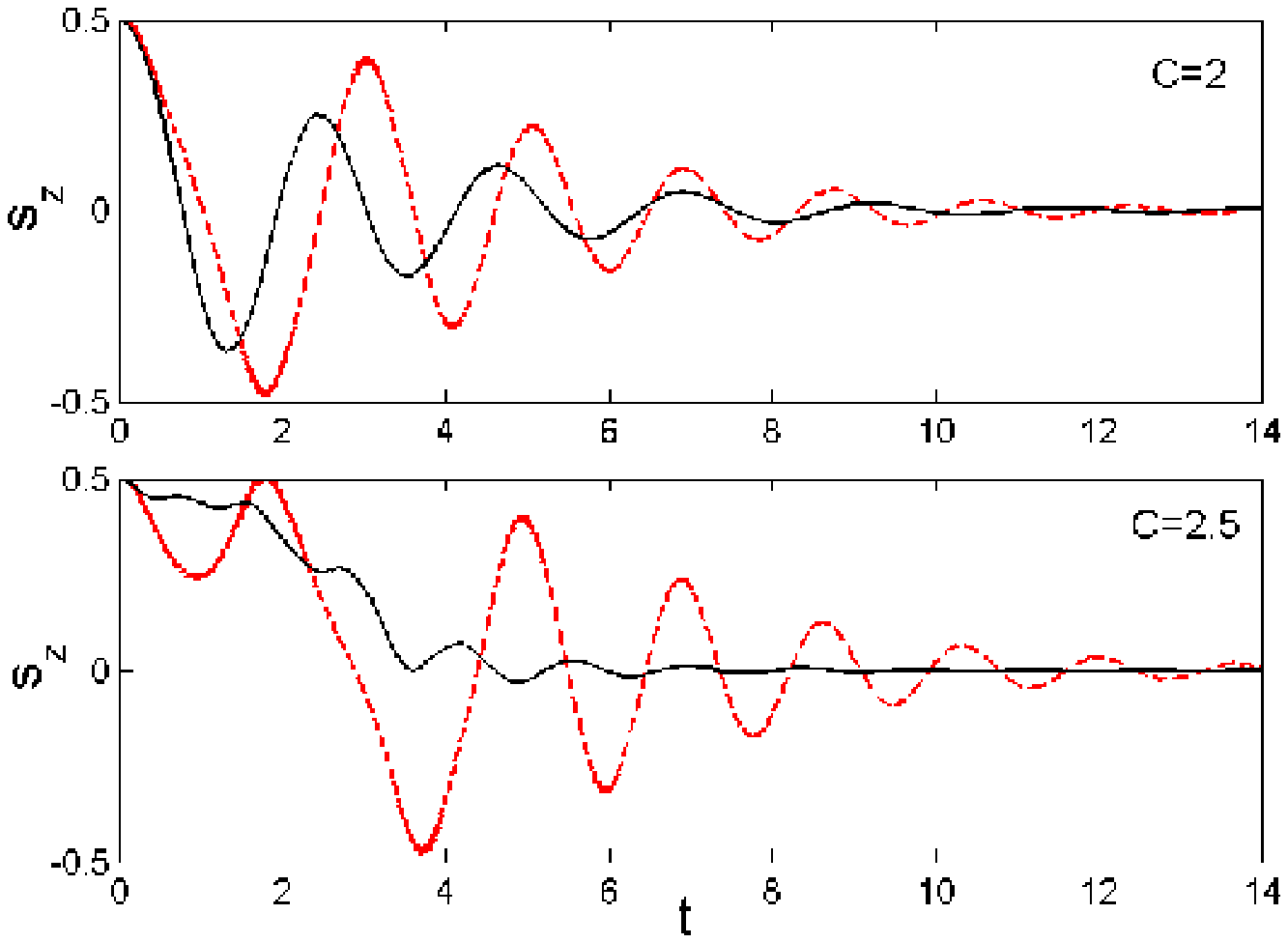}
\caption{(Color online) Time evolution of $s_z(t)$ in the case of $\gamma=0$ with $C=2$ and $C=2.5$. Here $\delta=0$, $\Omega = 1$, $\Lambda=0.2$, $\Upsilon=1$ and the initial state locates at the north pole. The dashed red lines and solid black lines correspond to mean-field and many-particle with $N=10$ behavior, respectively.}
\end{figure}

\begin{figure}[htp] \center
\includegraphics[width=3.5in]{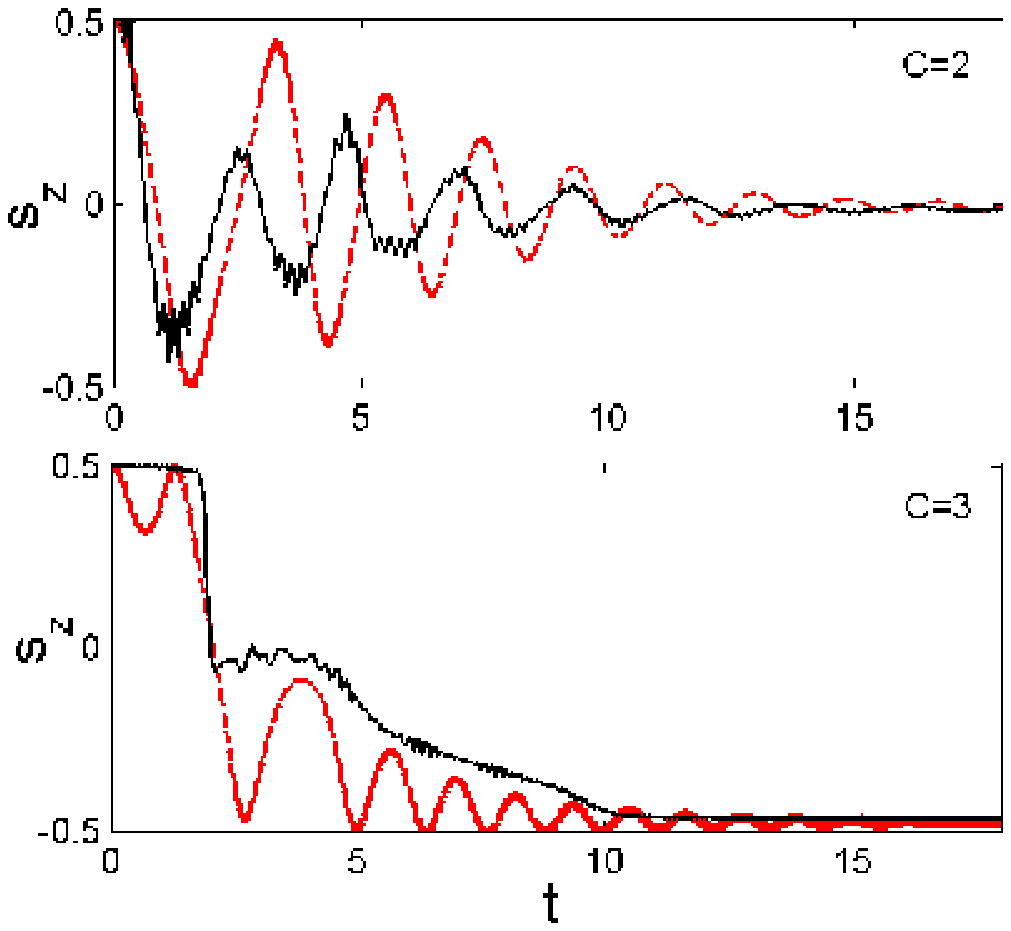}
\caption{(Color online) Time evolution of $s_z(t)$ in the case of $\gamma > \Lambda$ with $C=2$ and $C=3$. Here $\delta=0$, $\Omega = 1$, $\gamma=0.4$, $\Lambda=0.15$, $\Upsilon=1$ and the initial state locates at the north pole. The dashed red lines and solid black lines correspond to mean-field and many-particle with $N=10$ behavior, respectively.}
\end{figure}

\begin{figure}[htp] \center
\includegraphics[width=3.5in]{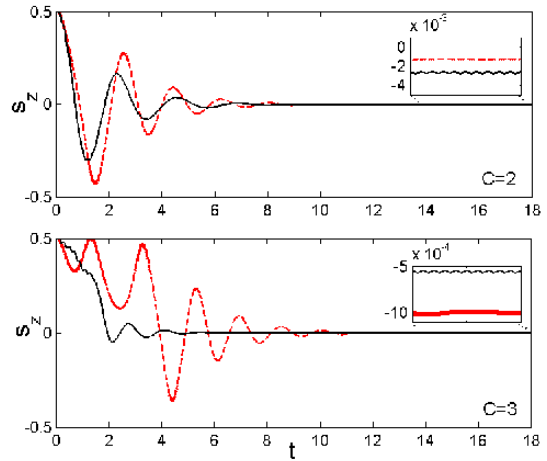}
\caption{(Color online) Time evolution of $s_z(t)$ in the case of $\gamma < \Lambda$ with $C=2$ and $C=3$. Here $\delta=0$, $\Omega = 1$, $\gamma=0.05$, $\Lambda=0.324$, $\Upsilon=1$ and the initial state locates at the north pole. The dashed red lines and solid black lines correspond to mean-field and many-particle with $N=10$ dynamics, respectively. The inset gives the enlarged region between the two dotted lines.}
\end{figure}

Below, we discuss how to control the population switching by controlling the nonlinearity and dissipation. We assume that all atoms initially occupy the second hyperfine level, which can be easily prepared in labs~\cite{zibold}. This initial state corresponds to the north pole of the Bloch sphere. In Fig. 5, we show the time evolution of $s_z(t)$ for the case of $\gamma=0$ with two different values of the nonlinearity. In this case, it is found that an initial self-trapped state can finally evolve into an equal-population state. In Fig. 6, we show the numerical results for $s_z(t)$ in the case of $\gamma>\Lambda$ with three different values of nonlinearity. From this numerical calculations, we find that the system may switch from an self-trapped state to another self-trapped state. The numerical data show that strong nonlinearity drives the system close to self-trapped state with large population imbalance, $s_{z}\approx-0.5$. In addition, for $\gamma<\Lambda$, the system will jump from an initial self-trapped state to an equal-population state, as illustrated in Fig. 7. The full many-particle result shows a very similar behavior. By varying the relative dissipation rate $\gamma$, one can observe the switching between different self-trapped states (see Fig. 6) or the switching from a self-trapped state to an equal-population state (see Figs. 5 and 7). On the contrary, one can also understand the dissipative mechanism of an open two-component BEC system by measuring the population difference between two hyperfine levels.

\section{Conclusions}

In summary, we explore the many-body quantum coherence and population dynamics in a two-component condensate coupled with a continuum. The system is described by a NH Bose-Hubbard dimer whose diagonal and off-diagonal Hamiltonian elements are both complex numbers. The combination of dissipation and nonlinearity may induce different steady states and modify the bifurcation of coherence. By tuning the nonlinearity and dissipation, the coherence enhancement exhibits. Particularly, different relative dissipation strengths between two hyperfine levels $\gamma$ will drive the two-component condensate into different stable states. Under the condition of $\gamma=0$, the atoms always evolve into a balanced state with equal population. In the case of $\gamma \neq 0$, dependent upon the values of $\gamma$ and $\Lambda$, the system will evolve into different final states. For $\gamma > \Lambda$, the system always evolve into a steady state with self-trapping. The change of nonlinearity and initial state not only can modify the evolution path and the dissipation velocity to the final state but also can change the strength of self-trapping. For $\gamma < \Lambda$, the atoms always evolve to the same quasi-equal-population state with a small $s_{z}$ in order of $10^{-3}$, although the evolution path might be different for different nonlinearity strengthes.

Our results show that the combination of dissipation and nonlinearity can be used to manipulate the steady behaviors, such as, controlling the transition between two self-trapped states or between a self-trapped state and a non-self-trapped state. Therefore, our results provide an alternative route for manipulating the many-body coherence and population dynamics by utilizing dissipation and nonlinearity.

With currently avaliable techniques, it is possible to realize our model in experiments.  It has suggested that autoionizing Rydberg states \cite{fano,zoller} can be used as a continuum. In recent, the coupling between Bose-Einstein condensed atoms and highly excited Rydberg states have been reported \cite{leslie}. Therefore, based on the experimental techniques for observing internal Josephson effects in a  two-component condensate~\cite{zibold}, by coupling the condensed atoms to a continuum of autoionizing Rydberg states, our results may be tested in future experiments.

\section*{Acknowledgments}

This work is supported by the NBRPC under Grants No. 2012CB821305, the NNSFC under Grants No. 11075223, 11147021, 10905019 and 11175064, the PCSIRT under Grant No. IRT0964, the Hunan Provincial Natural Science Foundation under Grant No. 12JJ4010, the NCETPC under Grant No. NCET-10-0850 and the
Ph.D. Programs Foundation of Ministry of Education of
China under Grant No. 20120171110022.

\end{document}